\documentstyle[aps,preprint]{revtex}
\begin{document}
\draft
\title{Polarizability and single particle spectra of 
two-dimensional s- and d-wave superconductors}
\author{Hae-Young Kee$^{1,2}$ and C. M. Varma$^2$}
\address{
${}^1$ Department of Physics, Rutgers University, Piscataway, NJ 08855-0849\\
${}^2$ Bell Laboratories, Lucent Technologies, Murray Hill, NJ 07974}
\maketitle

\begin{abstract}
We present analytic results for the polarizabilities, $\chi({\bf Q},\omega)$,
in the charge, spin, and current
channels for two-dimensional s- and d-wave superconductors
for large momentum transfer. 
A collective mode in the charge channel is predicted to exist 
for extremum vectors of the Fermi-surface  with energy 
below twice the maximum superconducting gap.
Such modes are directly observable through inelastic x-ray or electron
scattering. 
Scattering of single particle excitations by these collective modes
leads to several unusual features in the single particle spectrum in the
superconducting state which are seen in angle resolved photoemission 
experiments.

\end{abstract}

\section{Introduction}
It is now well established that the hole doped cuprates superconductors
are d-wave superconductors with a very strong two-dimensional anisotropy.
\cite{dwave}
It is clear that electron-electron interactions lead to the 
attractive pairing, although no consensus has emerged about the precise physical
nature of such interactions.
These interactions must also produce a very unusual spectra of particle-hole
fluctuations leading to the observed non-Fermi-liquid transport properties 
in the normal metallic state, at least, near the hole-density for
maximum $T_c$.

A phenomenological model\cite{cmv}
from which many of the anomalous transport properties can be
calculated consistently assumes that the particle-hole fluctuations are 
scale-invariant, as in the critical region of a quantum phase transition.
This model further assumes that the fluctuations are nearly momentum 
independent,
leading to the prediction that the single-particle scattering rate
-- measurable in angle resolved photoemission(ARPES) or tunneling spectra --
has the 
same anomalous temperature(and frequency) dependence as in the momentum 
scattering rate measured through the resistivity or the energy scattering rate
measured by thermal-conductivity. 
The prediction of an unusual single particle line shape was verified by
ARPES experiment.\cite{olson}

Any particle-hole spectra, usual or unusual, must develop a gap related
to the superconducting gap below $T_c$.
If the lifetime of quasi-particles in the normal state is primarily
determined by electron-electron interactions, it must 
become very long in the superconducting state
compared to the extrapolation from the normal state.\cite{kuroda}
This prediction has been also verified.\cite{nuss}
A further prediction followed from the realization that the particle-hole
spectra must revert to its normal state form at energies well above
twice the (maximum) superconducting gap.
Therefore quasi-particle 
line shape for energies above about $3\Delta$ must continue to
be of the normal state form.
The single-particle spectrum in the superconducting gap must then be a 
sharp peak near the BCS energy followed by a broad hump starting at about
$3\Delta$.\cite{littlewood}
These predictions have been also qualitatively verified through ARPES spectra.
\cite{shen,ding1,campuzano}

Recent more precise ARPES measurements\cite{ding2,norman}  have motivated us to examine more
precisely the change in the particle-hole spectra and the single
particle spectra in the superconducting state.
The ARPES spectra shows a slower variation with $\epsilon_{\bf k}$
of the sharp peak than BCS and a more robust hump near $3\Delta$
than expected. Also, the calculations must be performed using the 
d-wave anisotropy of the gap rather than the s-wave gap used in the
 simple calculation performed earlier\cite{littlewood}. 

To obtain the single particle spectra, we must calculate the polarizabilities
for the charge, current   and spin channels as a first step.
We expect that the momentum-dependence of the polarizability of d-wave superconductors
would be qualitatively different from that of s-wave superconductors.
The result for the polarizability of s-wave superconductors, at large momentum
for both two and three dimensions (and the associated anomalies in the phonon
spectra), were presented earlier.\cite{kee,kawano,stassis}
 We include the results for polarizability of s-wave superconductors,
to compare with d-wave superconductors.
For a spherical Fermi-surface in three dimensions, the lowest order
polarizability in charge channel has a logarithmic singularity at energy
$2\Delta$ for momentum close to the extremum vector 
spanning the Fermi-surface, $(2 k_F,0,0)$. 
In two dimensions the logarithmic
singularity is present near momentum $(k_F,k_F)$ but for momentum close to
$(2k_F,0)$, the singularity is stronger $\sim (\omega-2\Delta)^{-1/4}$.
We give further mathematical details in the Appendix A.

The singularities in the superconducting state
in the polarizabilities at large momentum, (while they are absent in the normal
state) arise because the singularity in the one particle density of states
in BCS-superconductors near $\Delta$ are strong enough to lead to singularities in the joint particle-hole densities of
states at special momentum-vectors connecting points on the Fermi-surface.
 In the d-wave case, these will be vectors connecting points on 
the Fermi-surface with the maximum gap. 
Taking these points to be along the x and the y-axes in two-dimensions, 
see Fig. 1, we can define two vectors  ${\bf Q}_0$ and ${\bf Q}_a$.

\begin{eqnarray}
\epsilon_{{\bf k}-\frac{{\bf Q}_0}{2}} 
 &=& -v_F k_x + \frac{k^2_y} {2 m_{\perp}},\nonumber\\
\epsilon_{{\bf k}-\frac{{\bf Q}_0}{2}+{\bf Q}_a} 
  &=& v_F k_y + \frac{k^2_x}{2 m_{\perp}}.
\label{Qa}
\end{eqnarray}
${\bf Q}_0$ will often be called the extremum vector. 
For an isotropic Fermi-surface, or more generally with 
a Fermi-surface of a square lattice we can interchange x and y 
in Eq.(\ref{Qa}).

We find that no nesting is required for the singularities. 
However nesting, i.e., $\frac{1}{m_{\perp}} = 0$ for points of the 
Fermi-surface (coincident with the maximum of the gap for the d-wave case )
leads to stronger particle-hole singularities.

The most important new result in this paper
is that the particle-hole spectra in 
two-dimensions has a singularity at energy below $2\Delta$ at large momentum
transfer in charge channel in d-wave 
superconductors leading to new collective modes.
Such collective modes can be observed directly by inelastic x-ray or electron
scattering. 
Thus the particle-hole spectra does not just develop a gap below $2\Delta$
as earlier assumed.\cite{littlewood}
These results are true even when the normal state is a Fermi-liquid.

The collective modes also strongly influence the single particle 
spectra measured by ARPES.
We compute the modified single-particle spectral function through 
the self energy near the maximum gap.
We find a sharp peak near $\Delta$ coming from the original BCS form
and the broad hump below $3\Delta$ arising from the exchange self energy due to the collective mode.

This paper is organized as follows.
We first define the model and show the general formula for
various response functions in section II.
We also discuss the effect of the coherence factors in this section.
We present analytical(and graphical) results of calculations of 
polarizabilities for both s-wave and d-wave superconductors,
in section III and IV, respectively.
In section IV, we also show single particle spectral function
and compare with the experiment.
In section V, we discuss the predictions, 
limitation of our results and 
comparison with existing experiments. 

\section{Model and Definitions}

We consider the simplest possible model for the electronic energy
\begin{equation}
\epsilon_{\bf k} =  \frac{k_x^2}{2m}
+ \frac{k_y^2}{2m}-\mu, \label{model}
\end{equation}
where $\mu$ is the chemical potential, $k_F^2/(2 m)$,
in a circular Fermi-surface of radius $k_F$ shown in Fig. 1.
Any model with nesting on the Fermi-surface will have more singular
polarizability than the isotropic surface, Eq. (\ref{model}).
We shall mention the expected corrections for such a case.

The single particle Green's function in the superconducting
state can be written in the Gorkov-Nambu notation as\cite{sch}
\begin{equation}
G({\bf k},\omega) = \frac{\omega I +\epsilon_{\bf k} \tau_3 +
\Delta_{\bf k} \tau_1}{\omega^2 - E_{\bf k}^2 + i\eta},
\end{equation}
where $E_{\bf k}^2 =(\epsilon^2_{\bf k}+\Delta_{\bf k}^2)$.
and  $\Delta_{\bf k}=\Delta$ for s-wave superconductor and
$\Delta_{\bf k}=\Delta \cos{(2\phi)}$ for d-wave superconductor.

The lowest order polarizability in charge, spin and current channels
at momentum ${\bf Q}$ and energy $\omega$ are given  at $T=0$ by
\begin{equation}
\chi_0^{charge}({\bf Q},\omega) =
 -i \int \frac{d^2 k d\omega^{\prime}}{(2 \pi)^3}
                Tr [\tau_3 G({\bf k}+{\bf Q},\omega+\omega^{\prime}) 
                   \tau_3 G({\bf k},\omega^{\prime})],
\label{chi0c}
\end{equation}

\begin{equation}
\chi_0^{spin}({\bf Q},\omega) =
 -i \int \frac{d^2 k d\omega^{\prime}}{(2 \pi)^3}
                Tr [ G({\bf k}+{\bf Q},\omega+\omega^{\prime}) 
                    G({\bf k},\omega^{\prime})].
\label{chi0s}
\end{equation}

\begin{equation}
\chi_{0,ij}^{current}({\bf Q},\omega) =
 -i\frac{e^2}{m^2} \int \frac{d^2 k d\omega^{\prime}}{(2 \pi)^3}
               ({\bf k}+{\bf Q}/2)_i ({\bf k}+{\bf Q}/2)_j
                Tr [ G({\bf k}+{\bf Q},\omega+\omega^{\prime}) 
                    G({\bf k},\omega^{\prime})].
\end{equation}
where $\tau_i$ are Pauli matrices.

After performing the frequency integral, 
they are given by
\begin{equation}
\chi_0^{charge}({\bf Q},\omega)=
-\frac{1}{2} \sum_{\bf k}
 \left( 1-\frac{\epsilon_{{\bf k}+{\bf Q}}\epsilon_{\bf k}
-\Delta_{{\bf k}+{\bf Q}} \Delta_{\bf k}}
{E_{{\bf k}+{\bf Q}}E_{\bf k}} \right)
\left( \frac{1}{\omega+E_{{\bf k}+{\bf Q}}+E_{\bf k}+i\eta}
- \frac{1}{\omega-E_{{\bf k}+{\bf Q}}-E_{\bf k}+i\eta} \right)
\label{charge}
\end{equation}

\begin{equation}
\chi_0^{spin}({\bf Q},\omega)=
-\frac{1}{2} \sum_{\bf k}
 \left( 1-\frac{\epsilon_{{\bf k}+{\bf Q}}\epsilon_{\bf k}
+\Delta_{{\bf k}+{\bf Q}} \Delta_{\bf k}}
{E_{{\bf k}+{\bf Q}}E_{\bf k}} \right)
\left( \frac{1}{\omega+E_{{\bf k}+{\bf Q}}+E_{\bf k}+i\eta}
- \frac{1}{\omega-E_{{\bf k}+{\bf Q}}-E_{\bf k}+i\eta} \right)
\end{equation}

\begin{eqnarray}
\chi_{0,ij}^{current}({\bf Q},\omega) & \propto &
\frac{1}{2} \sum_{\bf k} ({\bf k}+{\bf Q}/2)_i  ({\bf k}+{\bf Q}/2)_j
 \left( 1-\frac{\epsilon_{{\bf k}+{\bf Q}}\epsilon_{\bf k}
+\Delta_{{\bf k}+{\bf Q}} \Delta_{\bf k}}
{E_{{\bf k}+{\bf Q}}E_{\bf k}} \right) \nonumber\\
& & \left( \frac{1}{\omega+E_{{\bf k}+{\bf Q}}+E_{\bf k}+i\eta}
- \frac{1}{\omega-E_{{\bf k}+{\bf Q}}-E_{\bf k}+i\eta} \right)
\end{eqnarray}

We are interested in the response at large ${\bf Q}$, comparable in magnitude to
$k_F$. In such a case, in a lattice, the current response always has
a part proportional to the charge response.\cite{littlewood,abri}
This is missing in the translationally invariant model of Eq. (\ref{model}).

Let us discuss the importance of the coherence factors.
Note that the coherence factor contains
 $-\Delta_{{\bf k}+{\bf q}} \Delta_{\bf k}$ in the charge channel, while
 $+\Delta_{{\bf k}+{\bf q}} \Delta_{\bf k}$ in the spin channel.
Therefore in the case of s-wave superconductors, any interesting features near
Fermi-surface can arise only from the charge channel. 
However,  d-wave superconductors have a change of sign of the gap 
under $\pi/2$ rotation:  
\begin{equation}
\Delta_{\phi}=-\Delta_{\phi+\pi/2}.
\end{equation}
This leads to the result that the susceptibility 
at ${\bf Q}\sim (k_F,k_F)$ from
antinode(+) to antinode(-) is negligible
in the charge channel, while the susceptibility at ${\bf Q}\sim (2k_F,0)$ from
antinode(+) to antinode(+) is negligible 
in the spin channel.
Therefore, the interesting features in charge channel can arise from 
the excitation of ${\bf Q}\sim (2k_F,0)$ and that
in spin channel from the excitation of ${\bf Q}\sim (k_F,k_F)$ 
near $\omega =2\Delta$.
This issue is related to the sharp magnetic neutron scattering peak
which was observed in the superconducting state of $YBa_2Cu_3 O_7$
near momentum ${\bf Q}=(\pi,\pi,\pi)$ 
( which in our notation is $(k_F,k_F)$ and 
odd with respect to inversion between the two planes per unit cell), 
and energy $\omega =41 meV$.\cite{fong,mignod,mook}
It was pointed out in Ref. \cite{fong} that the BCS coherence factor
in neutron scattering amplitude vanishes unless $\Delta_{\bf k}$ and 
$\Delta_{\bf k+Q}$ have opposite signs, because magnetic scattering
is odd with respect to the time reversal symmetry.
This was argued in Refs. 
\cite{fong,bulut,monthoux,maki,stemmann,lavagna,onuf}  to show that
the peak is a manifestation of the $d_{x^2-y^2}$ state.
Another suggestion was that the superconducting order parameter has
s-wave symmetry and opposite signs in the bonding and antibonding
electron bands formed with a $Cu_2O_4$ bilayer.\cite{yakovenko}
A numerical calculation shows a collective antiferromagnetic
excitation for momentum $(\pi,\pi,\pi)$ with  d-wave gap.\cite{sca}. 
These calculations concentrated only on the response function 
in the spin channel.

We are interested in the polarizability for the momentum vectors 
connecting the maxima of the absolute value of the gap.
Near such a maximum gap, we can expand the gap
in momentum space, e.g.,
\begin{eqnarray}
\Delta_{{\bf k}+{\bf Q}_0/2}
& =& \Delta (\cos^2{\phi}-\sin^2{\phi}) \nonumber\\
&=& \Delta \frac{(k_F+k_x)^2-k_y^2}{(k_F+k_x)^2+k_y^2}
 \approx  \Delta (1-2\frac{k_y^2}{k_F^2}).
\label{dgap}
\end{eqnarray} 
 ${\bf Q}_0$ connects the (+) antinode 
and (+) antinode and ${\bf Q}_a$ connects the (+) antinode
and (-) antinode as shown in Fig. 2.
The dispersion relations, e.g., Eq. (\ref{Qa}),
show that one direction of
the dispersion is more slowly varying than the other;
it is linear in $k_x(k_y)$ but quadratic in $k_y(k_x)$.
At a general ${\bf Q}$, the  dispersion is linear
in both $k_x$ and $k_y$.
A singular behavior of polarizability, if it exists, is therefore more 
likely for momentum near ${\bf Q}_0$ and ${\bf Q}_a$ 
at frequencies near $2\Delta$.

\section{s-wave superconductors}

The long-wavelength collective modes for s-wave superconductors have
been discussed earlier.\cite{littlewood,anderson,nambu}
We concentrate on the results for large momenta.
The results for the charge polarizability in this case have been
already presented.\cite{kee}
The details of the derivation are given in Appendix A.
Near the extremum vector, ${\bf Q}_0$, in two dimensions, we found 
\begin{eqnarray}
{\rm Re} \chi_0^{charge}({\bf Q},\omega) &\simeq&
-\frac{\Delta^{3/4} N_0}
{E_F^{1/2} } 
\hspace{1mm} \frac{1}{|\delta \omega|^{1/4}}
\nonumber\\
{\rm Im} \chi_0^{charge}({\bf Q},\omega)& \simeq & 
\cases { 0 & $ \delta \omega < 0
$\cr
\frac{\pi}{2} \frac{\Delta^{3/4}N_0}
{E_F^{1/2}} \hspace{1mm} \frac{1}{|\delta \omega|^{1/4}}
&  $\delta \omega > 0 $, \cr}
\end{eqnarray}
where $\delta \omega \equiv \omega - 2 \Delta(1+\xi^2 q^2/4)$,
and $N_0 = v \frac{m}{2 \pi}$ with $v$, the area
of unit cell. (from now on, set $v=1$.)
Here,  $q$ is defined to be parallel component of 
$({\bf Q}-{\bf Q}_0)$ satisfies $q \xi << 1$,
where $\xi$ is the coherence length of the superconducting gap,
$\sim \hbar v_F/\Delta$.

Another singularity has been also found 
for momentum ${\bf Q}_a=(k_F,k_F)$
at energy $2\Delta$ in agreement with earlier results(see, e.g.,
\cite{yakovenko}).
\begin{eqnarray}
{\rm Re} \chi_0^{charge}({\bf Q}_a,\omega) &\simeq&
-\frac{N_0 \Delta}
{E_F } \hspace{1mm}  {\rm ln}\frac{E_F}{\sqrt{|\delta \omega|
\Delta}}
\nonumber\\
{\rm Im} \chi_0^{charge}({\bf Q}_a,\omega)& \simeq & 
\cases { 0 & $ \delta \omega < 0 $\cr
\frac{\pi}{2} \frac{ N_0 \Delta}{E_F}
&  $\delta \omega > 0 $, \cr}
\end{eqnarray}
where $\delta \omega =\omega-2\Delta$.

For completeness, we also give the results for three dimensions.
Singular behavior is also found in three dimensions, but it is less
singular than that of two dimensions.
\begin{eqnarray}
{\rm Re} \chi_0^{charge}({\bf Q}_0,\omega) &\simeq&
-\frac{N_0 \Delta}
{E_F } \hspace{1mm}  {\rm ln}\frac{E_F}{\sqrt{|\delta \omega|
\Delta}}
\nonumber\\
{\rm Im} \chi_0^{charge}({\bf Q}_0,\omega)& \simeq & \cases { 0 & $ \delta \omega < 0
$\cr
\frac{\pi}{2} \frac{ N_0 \Delta}
{E_F}
&  $\delta \omega > 0 $, \cr}
\label{swave}
\end{eqnarray}
where 
$N_0 = \frac{\sqrt{m^3}}
{2 \pi^2}$ 
The singularity in the charge channel couple to phonons.
Eq. (\ref{swave}) has been used to understand the anomalies generated
in the lattice vibration spectra in the superconducting state in some compounds.
\cite{kee,kawano,stassis}
It is clear that for nested Fermi-surfaces, the polarizability become
more singular.
We found that the nesting along the y-direction for ${\bf Q}_0$
gives $1/\sqrt{|\omega-2\Delta|}$ in the polarizability. 

\section{d-wave superconductors}
The calculation of the polarizability in d-wave superconductors is more complicated than for s-wave superconductors.
Details of the calculation are given in Appendix. B. We present the results below. 

We again concentrate on the case of large momentum transfers, near  ${\bf Q}_0$ and ${\bf Q}_a$. Referring to Fig. 2, we find that near ${\bf Q}_0$, at
 ${\bf Q}=(2k_F-q_x,q_y)$ for $|q_y| < \sqrt{\frac{\Delta}{E_F}}$
and $|q_x| < (\frac{\Delta}{E_F})$, where $k_F =1$, the lowest order charge susceptibility is given by
\begin{eqnarray}
{\rm Re} \chi_0^{charge}({\bf Q},\omega) &\simeq&  \cases{
- A {\rm ln}\left( \frac{ \sqrt{8\Delta}}
{\sqrt{\delta\omega-4\Delta^3/E_F^2}} \right), &  
 $4\frac{\Delta^3}{E_F^2} < \delta \omega $  \cr
- B {\bf K} (X),
&
$ -\Delta q_y^2-\frac{E_F^2 q_y^4}{16 \Delta}
< \delta\omega < 4\frac{\Delta^3}{E_F^2}$  \cr
-B/2,    &
$ \delta \omega <  -\Delta q_y^2-\frac{E_F^2 q_y^4}{16 \Delta}$, \cr }
\label{chargere}
\end{eqnarray} 

\begin{eqnarray}
{\rm Im} \chi_0^{charge}({\bf Q},\omega) &\simeq&  \cases {
 0, & 
$ 4\frac{\Delta^3}{E_F^2} < \delta\omega $ \cr
- \frac{B}{2} {\rm ln} \left( \frac{X N_0}{ 4 B} \right),
&
$ -\Delta q_y^2-\frac{E_F^2 q_y^4}{16 \Delta}
< \delta\omega < 4\frac{\Delta^3}{E_F^2}$  \cr
- \frac{B}{2} {\rm ln} \left( X \right), & 
$ \delta \omega <  -\Delta q_y^2-\frac{E_F^2 q_y^4}{16 \Delta}$, \cr }
\label{chargeim}
\end{eqnarray} 
where 
\begin{eqnarray}
X &= & \left( \frac{
\sqrt{|1+(E_F^2 q_y^2)/(8\Delta^2)-\sqrt{1-\delta\omega E^2_F/(4\Delta^3)}|}}
{\sqrt{1+(E_F^2 q_y^2)/(8\Delta^2)+\sqrt{1-\delta\omega E^2_F/(4\Delta^3)}}}
\right),  \nonumber\\
A&=& \frac{N_0}{4\sqrt{2}\sqrt{1+(E_F^2 q_y^2)/(8 \Delta^2)}},\nonumber\\
B&=& \frac{N_0}{2\sqrt{2}\sqrt{1+(E_F^2 q_y^2)/(8\Delta^2)+
\sqrt{1-\delta\omega E^2_F/(4\Delta^3)}}}, 
\end{eqnarray}
and $\delta \omega=2\Delta-2\Delta q_y^2-\omega$.
${\bf K}(X)$ is the Elliptic integral of first kind. 
The results above show that there is a logarithmic singularity 
at $\omega=2\Delta-2\Delta q_y^2-4\Delta^3/E_F^2$  in 
${\rm Re} \chi^{charge}_0$
and $\omega=2\Delta-\Delta q_y^2+(E_F^2 q_y^4)/(16 \Delta)$
 in ${\rm Im} \chi^{charge}_0$.
The susceptibility is plotted in Fig. 3.
The singularity arises just below $2\Delta$. The shift from $2\Delta$, 
of order $\Delta^2/E_F^2$, comes from the curvature of the gap.
In the case of s-wave superconductor\cite{kee}, the singular behavior
of $|\delta \omega|^{-1/4}$ at ${\bf Q}=(2k_F,0)$ remains
at $\omega =2\Delta$.
For d-wave superconduictors, the momentum dependence of the gap does not
wipe out the singularity, but reduces it to logarithmic and
changes the position of the singularity.

As explained earlier, the singularity in the polarizability in the spin channel can arise if the momentum transfer
connects opposite signs of the gap.
The susceptibility in the spin channel near ${\bf Q}_a$, at  
 ${\bf Q}=(k_F-q_x,k_F-q_y)$ for $|q_x| < \frac{\Delta}{E_F}$ 
and $|q_y| < \frac{\Delta}{E_F}$, is given by

\begin{eqnarray}
{\rm Re} \chi_0^{spin}({\bf Q},\omega) &\simeq&
-\frac{N(0) \Delta}
{E_F \sqrt{1+q_x^2} } 
 {\rm ln}\frac{E_F}{\sqrt{|\delta \omega| \Delta}}
\nonumber\\
{\rm Im} \chi_0^{spin}({\bf Q},\omega)& \simeq & \cases { 0 & $ \delta \omega < 0
$\cr
\frac{\pi}{2} \frac{ N(0) \Delta}
{E_F \sqrt{1+q_x^2}}
&  $\delta \omega > 0 $, \cr}
\end{eqnarray}
where $\delta\omega=2\Delta-2\Delta q_x^2-8\Delta^3 q_x^2/E_F^2-\omega$.

As we see, there is also a logarithmic singularity in the spin channel. 
But it is important  to note that the amplitude of the singularity
of the charge channel is bigger than that of the spin channel by the 
 $O(E_F/\Delta)$.

\subsection{Susceptibility in the Random Phase Approximation}

The next step is to see how the lowest order polarizability is modified
due to the interactions among the quasiparticles.
We present the approximate renormalized polarizability
in the present section.
In the random phase 
approximation,
\begin{equation}
\chi=\frac{\chi_0}{1+V \chi_0} 
\end{equation}
where we assume  that the interaction $V_{Q}=-V$. 
This has a singular part in charge channel
with a pole at frequency $\omega_Q$ with weight $r_Q$ and 
a regular part which
is approximately a constant for $\omega > 1.5 \Delta$.
The position of the pole, $\omega_Q$, in $\chi$ is given by 
\begin{eqnarray}
\omega_Q &=& 2\Delta-2\Delta q_y^2 -4\Delta^3/E_F^2 - 8\Delta e^{-\frac{1}{AV}}
\nonumber\\
&=& 2\Delta-2\Delta q_y^2-4 \Delta^3/E_F^2 -8\Delta 
e^{-\frac{\sqrt{8+(E_F^2 q_y^2)/\Delta^2}}{N_0 V}}.
\end{eqnarray}
The weight of the pole is
\begin{eqnarray}
r_Q&=&\frac{8\Delta}{AV^2} e^{-\frac{1}{AV}}\nonumber\\
&=& \frac{16\sqrt{2}\Delta \sqrt{1+(E_F^2 q_y^2)/(8 \Delta^2)}}{N_0 V^2}
e^{-\frac{\sqrt{8+(E_F^2 q_y^2)/\Delta^2}}{N_0 V}}.
\end{eqnarray}
Four parameters, $N_0 V$,
$\Delta/E_F$, $\Delta/V$, and $q_y$ determine the position and the strength of the poles.
Let us fix $N_0 V \sim 1$ and $\Delta/E_F \sim \Delta/V \sim 0.1$.
At $q_y=0$, the position of pole is at $\omega= 1.49 \Delta$
and the weight is $r=0.13$.
At $q_y=0.2$, the position of pole is at $\omega= 1.62 \Delta$
and the weight is $r=0.087$.
For the reasonable condition $|q_y| < \sqrt{\Delta/E_F} \sim 0.3$, the pole is at  about $\omega_0=1.5\Delta$
and the weight is $\sim 0.1$.

\subsection{Polarizability with a marginal Fermi-liquid vertex}

The anomalous normal state transport properties near the doping for 
maximum $T_c$ can be understood through a phenomenological
scale-invariant particle-hole spectrum, (Eq. (\ref{mfl}), see below)
 which at frequencies large
compared to temperature is a constant, independent of momentum(except in
the hydrodynamic regime) and of frequency up to a cut-off $\nu_c$.
\begin{equation}
{\rm Im} \chi_{MFL} (q,\nu) \approx N(0) ;\;\;\; T << \nu < \nu_c,
\label{mfl}
\end{equation}
The polarizability can in general be represented as in Fig. 4.
If we take for $\Gamma$ a phenomenological, momentum independent vertex
whose imaginary part is 
$\approx i N(0)^{-1}$ for $T << \nu < \nu_c$,
\begin{equation}
\chi_{MFL}(q,\nu) = \Gamma(\nu) \left( \sum_{k,\omega} \Lambda_{k,q}
G(k+q,\omega+\nu) G(k,\omega) \right)^2,
\label{mfl2}
\end{equation}
Eq. (\ref{mfl}) is then essentially reproduced.
Here $G$ is the normal state Green's function and $\Lambda$'s are
appropriate vertices.
In the preceding section, we put $\Gamma \sim \delta_{q,q^{\prime}}
\delta_{\omega,\omega^{\prime}}$ in Fig. 4. 
This leads to major difference between $\chi_{MFL}$ in the normal state
and $\chi_0$ of Eqs. (\ref{chi0c}) and (\ref{chi0s}).
But for the superconducting state, the results are qualitatively similar
if we make the further assumption that $\Gamma(\nu)$ does not change in the
passage to the superconducting state. 
This assumption is almost certainly incorrect.
But as long as $\Gamma(\nu)$ does not develop any sharp features for
energies below $2\Delta$ in the superconducting state, the results are
unlikely to be qualitatively altered. 
This is because of the sharp drop in $\chi_0$ for energies below
$2\Delta$ which we have calculated.
Using (\ref{chargere}) in (\ref{mfl2}), we see that the $log$
singularities near $2\Delta$ change to $log^2(\cdots)$ singularities.
Within the approximation of the phenomenology this difference is not important. 
There is one important difference however, $\chi$ in Fig. 3 and in Eq.
(\ref{mfl2}) is the total polarizability, not the reducible polarizability.
So no further manipulation as in section (A) is now not permitted.

\subsection{Single particle spectral function in the superconducting
state}
The singularity in the polarizability at large momentum provides
important modification in the single particle spectra.
The singular contribution to the polarizability in the spin channel 
and in the current channel for translationally invariant case is
$O(\Delta/E_F)$ smaller than the charge channel.
As we already mentioned, in a lattice the polarizability
in the current channel is proportional (and usually of similar magnitude)
to the charge channel.
We may therefore ignore the spin channel.

Generalized self energy can be obtained using the diagram shown
in Fig. 5.\cite{sch}
\begin{eqnarray}
\Sigma(k) &=& i \int \frac{d^3 q}{(2\pi)^3}
(V \tau_3) \psi^{\dagger}_{k-q} \chi(q) \psi_{k-q} (V \tau_3 )
\nonumber\\
&=& i\int \frac{d^3 q}{(2\pi)^3} V^2 \tau_3 G(k-q) \chi(q) \tau_3
\end{eqnarray}
where $k=({\bf k},\omega)$.
Here we neglect the other self energy contribution shown in Fig. 5(c) ;
\begin{equation}
 -i\int \frac{d^3 q}{(2\pi)^3} V^2 \tau_3  
Tr[\tau_3 G(k-q)]
\end{equation}
It is clear that this is negligible because it is of the order of
$O(\epsilon_{\bf k}/E_{\bf k})$ and
 we are near the Fermi-surface.

We write $\chi({\bf Q},\omega)$ as the sum of a singular part and
a regular part.
\begin{equation}
\chi({\bf Q},\omega) \approx \frac{2\omega_Q
r_Q}{\omega^2-\omega_Q^2+i\eta} f(Q) + \chi_{reg}.
\end{equation}
We will define $\Sigma^{sing}$ as the contribution of the first part
and $\Sigma^{reg}$ as the second one with 
\begin{equation}
\Sigma(\omega,{\bf k})= 
\Sigma^{sing}(\omega,{\bf k}) + \Sigma^{reg}(\omega,{\bf k}).
\end{equation}

For a d-wave superconductors $f(Q)$ gives the momentum limits shown
as above, ${\bf Q}=(2k_F-q_x,q_y)$
where $|q_x| < \Delta/E_F$ and  $|q_y| < \sqrt{\Delta/E_F}$ setting
$k_F=1$.
We assume that $\omega_Q=\omega_0$, and $r_Q=r$ in this range.

The self energy, $\Sigma^{sing}(\omega, {\bf k})$
 in the charge channel is
\begin{eqnarray}
\Sigma^{sing}(\omega,{\bf k})& =&i\int \frac{d\omega_1 d^2Q}{(2\pi)^2}
(V \tau_3) G_0(\omega+\omega_1,{\bf k+Q})(V \tau_3) \chi(\omega_1,{\bf Q})\nonumber\\
&=&i\int \frac{d\omega_1 d^2Q}{(2\pi)^2} (V\tau_3)
\frac{(\omega+\omega_1) I +\epsilon_{{\bf k}+{\bf Q}} \tau_3+
\Delta_{{\bf k}+{\bf Q}} \tau_1}
{(\omega+\omega_1)^2-E_{{\bf k}+{\bf Q}}^2+i\eta} (V\tau_3)
\frac{2\omega_0 r}{\omega_1^2-\omega_0^2+i\eta}.
\end{eqnarray}
After performing the frequency intergral, we found that the singular
part of the self energy is given by
\begin{eqnarray}
\Sigma_I(\omega,{\bf k}) & =& V^2 r \int\frac{d^2Q}{8\pi^2} (\frac{1}{2}) (\frac{1}
{\omega+\omega_0+E_{{\bf k}+{\bf Q}}+i\eta}
 \nonumber\\
& &  +\frac{1}{\omega-\omega_0-E_{{\bf k}+{\bf Q}}+i\eta})
\end{eqnarray}

\begin{eqnarray}
\Sigma_{\tau_1}(\omega,{\bf k}) &=&  V^2 r\int \frac{d^2Q }{8\pi^2} 
(\frac{\Delta_{{\bf k}+{\bf Q}}}{2E_{{\bf k}+{\bf Q}}})
(\frac{1}{\omega+\omega_0+E_{{\bf k}+{\bf Q}}+i\eta}  \nonumber\\
& &-\frac{1}{\omega-\omega_0-E_{{\bf k}+{\bf Q}}+i\eta} ),
\end{eqnarray}

\begin{eqnarray}
\Sigma_{\tau_3}(\omega,{\bf k}) & = -& V^2 r\int\frac{d^2Q}{8\pi^2}
(\frac{\epsilon_{{\bf k}+{\bf Q}}}{E_{{\bf k}+{\bf Q}}})
(\frac{1}
{\omega+\omega_0+E_{{\bf k}+{\bf Q}}+i\eta}
 \nonumber\\
& &  +\frac{1}{\omega-\omega_0-E_{{\bf k}+{\bf Q}}+i\eta}).
\end{eqnarray}

After expanding $E_{{\bf k}+{\bf Q}}$ near maximum gap(+),  we obtained
the self energy for $k_x \sim -k_F+k_x^{\prime}$ where $|k_x^{\prime}|
< |q_x|$ and $|k_y| < |q_y|$,
\begin{eqnarray}
{\rm Re}\Sigma_I(\omega,{\bf k}) &= & \frac{V^2N_0 r}{4\sqrt{2}} [
\arcsin{\frac{2\Delta}{\sqrt{E_F(\omega^*-\omega)}}}
\theta(-\omega+\omega^*-\frac{4\Delta^2}{E_F})\nonumber\\
& & -\arcsin{\frac{2\Delta}{\sqrt{E_F(\omega^*+\omega)}}}
\theta(\omega+\omega^*-\frac{4\Delta^2}{E_F})  ] \nonumber\\
{\rm Im}\Sigma_I(\omega,{\bf k}) &=& \frac{V^2N_0 r}{4\sqrt{2}} [
-{\rm arcsinh}{\frac{2\Delta}{\sqrt{E_F(-\omega^*+\omega)}}}
\theta(\omega-\omega^*)\nonumber\\
& &+{\rm arcsinh}{\frac{2\Delta}{\sqrt{E_F(-\omega^*-\omega)}}}
\theta(-\omega-\omega^*)  ]
\label{selfre}
\end{eqnarray}

\begin{eqnarray}
{\rm Re}\Sigma_{\tau_1}(\omega,{\bf k}) &= & -\frac{V^2N_0 r}{4\sqrt{2}} [
\arcsin {\frac{2\Delta}{\sqrt{E_F(\omega^*-\omega)}}}
\theta(-\omega+\omega^*-\frac{4\Delta^2}{E_F})\nonumber\\
& & +\arcsin {\frac{2\Delta}{\sqrt{E_F(\omega^*+\omega)}}}
\theta(\omega+\omega^*-\frac{4\Delta^2}{E_F})  ] \nonumber\\
{\rm Im}\Sigma_{\tau_1}(\omega,{\bf k}) &=& \frac{V^2N_0 r}{4\sqrt{2}} [
{\rm arcsinh}{\frac{2\Delta}{\sqrt{E_F(-\omega^*+\omega)}}}
\theta(\omega-\omega^*)\nonumber\\
& & +{\rm arcsinh}{\frac{2\Delta}{\sqrt{E_F(-\omega^*-\omega)}}}
\theta(-\omega-\omega^*)  ],
\label{selfim}
\end{eqnarray}
where $\omega^*=\omega_0+\Delta-4\Delta k_y^2+E_F^2/\Delta
(k_y^2-2 k^{\prime}_x)^2$.
$\Sigma_{\tau_3}$ is basically the same as 
$\epsilon_k/\Delta_k \Sigma_{\tau_1}$,
expanding the $\epsilon_{{\bf k}+{\bf Q}}=-\epsilon_{\bf k}+O({\bf q})$ 
with $E_{{\bf k}+{\bf Q}}=E_{\bf k}$.

We now discuss $\chi_{reg}$. 
This depends on whether one uses the form Eqs. (\ref{chargere}) and
(\ref{chargeim}) for $\chi_0$ or the marginal form.
This difference is important for large frequencies, because
 the single particle spectra for the former case reverts to
 the Fermi liquid form, which disagrees with experiments. 
So we consider the marginal fermi-liquid form. 
An important issue in that case is the form of $\chi_{reg}$
for low frequencies in the d-wave superconductor. 
In a d-wave superconductor, the regular part of $\chi_{reg}({\bf Q},\nu)$ 
at low energies is dominated by the particle-hole excitations 
with ${\bf Q}$ connecting the nodes of the gap function $\Delta_{\bf k}$. 
These determine the lifetime of low energy quasiparticles and 
the low temperature thermodynamics. 
But they are not involved in the scattering of quasiparticles 
near the maximum of the gap due to restrictions of momentum conservation. 
For the relaxation of such quasiparticles, 
we may therefore assume a marginal Fermi-liquid form 
for energies above $2\Delta$:
\begin{equation}
\chi_{reg} ({\bf Q},\nu) =r_{reg}({\bf Q}) N_0 \theta(\nu-2\Delta) , \;\;\; \nu < \nu_c,
\label{reg}
\end{equation}
with $r_{reg}+r =1$.
This produces
\begin{equation}
{\rm Im}\Sigma^{reg}_{I(\tau_1)} \propto \mp |\omega| \theta(\omega-3\Delta)
+  |\omega| \theta(-\omega-3\Delta).
\end{equation}

Fig. 6 shows the Re and Im parts of self-energy for excitations near 
the maximum gap including both the singular and the regular parts of the fluctuations.

The one particle Green's function can be obtained using
\begin{eqnarray}
G^{-1}&=& G_0^{-1}-\Sigma \nonumber\\
& =&  \left( \begin{array}{cc} 
\omega-\epsilon_{\bf k}+\Sigma_{\tau_3}-\Sigma_I & 
 -\Delta_{\bf k}-\Sigma_{\tau_1}\\
  -\Delta_{\bf k}-\Sigma_{\tau_1} & 
\omega+\epsilon_{\bf k}-\Sigma_{\tau_3}-\Sigma_I 
\end{array}
\right)
\end{eqnarray}
The spectral function is give by:
\begin{eqnarray}
A(\omega,{\bf k})&=& 
\frac{-sgn(\omega)}{\pi}
 {\rm Im} G_{11} \nonumber\\
&=& \frac{-sgn(\omega)}{\pi}
{\rm Im} \frac{\omega-\epsilon_{\bf k}+\Sigma_{\tau_3}-\Sigma_I}
{(\omega-\epsilon_{\bf k}+\Sigma_{\tau_3}-\Sigma_I)
(\omega+\epsilon_{\bf k}-\Sigma_{\tau_3}-\Sigma_I)
-(\Delta_{\bf k}+\Sigma_{\tau_1})^2 +i\eta}. 
\end{eqnarray}
The spectral function is plotted in Fig. 7 for ${\bf k}$ 
near  the maximum point of the gap and for the representative values, $V \sim E_F = 10 \Delta$. 
For  $ 0 <\omega < 2\Delta$ where 
${\rm Im} \Sigma_I={\rm Im} \Sigma_{\tau_1}=0$, it is 
a delta function near $\omega \sim \Delta$ and has a broad hump at $\omega \sim 2.5\Delta$.
As we change ${\bf k}$, we find the peak near $\Delta$ shifts
slower than the change of $\epsilon_{\bf k}$, itself.
The position of the broad hump shifts by  $ \Delta k_y^2$ which can be
up to order of $\Delta^2/E_F$. 
The broad peak would become smooth if the contributions from the
other excitations, e.g., node-node excitations are included.
The excitations  from node to node contribute to finite 
life time of quasiparticle for $\omega > \Delta$ which are not
included here. However, this would not change the qualitative behavior
of the broad peak. 

\section{discussion}
We have found a new collective mode  
for large momentum at energy just below $2\Delta$,
which can be detected by inelastic x-ray scattering or electron scattering.
Since neutrons have a negligible cross-section for scattering 
with electronic density fluctuations, 
this mode cannot be detected by neutron scattering. 
The calculations, done using an isotropic Fermi-surface give a lower 
limit to the spectral weight of the collective modes. 
Realistic band-structures show some
nesting in the region of the maximum gap due to which stronger effects 
are to be expected in experiments.

We have also calculated the single particle spectral function which 
is strongly affected for states near the maximum of the gap by the 
collective mode. This quantity has been measured by ARPES
\cite{norman}.
ARPES on $Bi2212$ near $(\pi,0)$
in the superconducting state shows two features.
There exists a sharp peak around $40 meV$ and a broad hump at about $100 meV$.
Our results also show two peaks in the spectral function, a sharp peak 
near $\Delta$
and a broad peak near $2.5\Delta$.
When we change the momentum by {\bf k} in a direction perpendicular 
to the Fermi-surface from the point of gap-maximum, the sharp peak  
and the broad peak shift as $E_F k^{\prime}_x$ where 
$k^{\prime}_x$ can vary up to $\Delta/E_F$ (See Fig. 7 (d), and
Fig. 2 (c) in Ref. \cite{norman}). 
For a change of momentum transverse to the Fermi-surface, 
this variation is cancelled
to leading order by a similar decrease in the value of the gap. 
(See Fig. 7 (b), (c), and Fig. 2 (b) in Ref. \cite{norman})
These features are generally consistent with the experiments, 
although the persistence of the sharp feature near $\Delta$ is over 
a wider region than calculated by us. We think this
is probably due to the nesting and Van-Hove dispersions of 
the band-structure near the points of gap-maximum.

It is also apparent in the experiments that
when the sharp peak disappears, the broad peak start to move rapidly
as in the normal state. This is also consistent with our expectation that the
behavior of
spectral function must be similar to that of normal state for $\epsilon_{\bf k}$
 away from the chemical potential by much more than $\Delta$.

We, finally,  mention the limitation of our results. 
Due to the use of an isotropic Fermi-surface,
the range of momentum where our results are valid
is rather limited.
Another limitation of our results is that 
the  one particle Green function $ G $ is not calculated self-consistently. 
We start with $ G $ of the  BCS form.
The singular self energy corrections split the BCS peak 
in the spectral function
into two peaks. 
But we do not go back and calculate the particle-hole fluctuations 
and the self-energy with the renormalized $G$.
We expect that using renormalized  particle $G$, 
additional features  at higher energy( near $5\Delta$) 
appears and the features already calculated are sharpened. 
The magnitude of these higher order effects is expected to be of order
 $N(0)V\Delta/E_F$. 
We have not included the vertex corrections in a systematic fashion 
either but have contented ourselves with a phenomenological 
requirement that the vertex be such that the spectral 
function in the superconducting state reverts to that of the normal 
state at large enough energy.
Another limitation is the assumption of the sharp change as a function
of $\nu$ in Eq. (\ref{reg}).
We expect a rapidly varying but smooth function.
This would tend to broaden the onset of the hump in the spectral functions in 
Fig. 7. 

\appendix
\section{}
We show how to evaluate the integrals to obtain the results quoted in Ref.
\cite{kee} and section III of this paper. 
After performing the frequency integral in Eq. (\ref{chi0c}), one must 
evaluate
\begin{equation}
\int d^{d} k  \frac{E_{\bf k} E_{{\bf k}+{\bf Q}}-\epsilon_{\bf k}
\epsilon_{{\bf k}+{\bf Q}}+\Delta^2}{E_{\bf k} E_{{\bf k}+{\bf Q}}} 
(\frac{1}{ E_{\bf k}+E_{{\bf k}+{\bf Q}}-\omega+i\eta}+ 
\frac{1}{ E_{\bf k}+ E_{{\bf k}+{\bf Q}} +\omega-i\eta})
\label{int1}
\end{equation}
The coherence factor does not play an important role for large ${\bf Q}$
 so that we simply put it $O(1)$.
Both $E_{\bf k}$ and $E_{{\bf k}+{\bf Q}}$ are bounded from below by
$\Delta$. 
Now expand $E_{\bf k}$ and $E_{{\bf k}+{\bf Q}}$ near the chemical potential
in cartesian coordinates and (without loss of generality) take ${\bf Q}$ along
$x$-axis.
First consider ${\bf Q}_0= 2 k_F \hat{x}$, 

\begin{equation}
\epsilon_{\bf k} = -v_F p_x+\frac{p_y^2}{2m}+ \frac{p_z^2}{2m}+O(p_x^2)
\label{exp1}
\end{equation}
where $p_x = (k_x-k_{Fx})$, $p_y = (k_y-k_{Fy})$ and $p_z =(k_z-k_{Fz})$,
$p$'s are $\ll k_F$'s, where $k_F=(k_{Fx}, k_{Fy}, k_{Fz})$.
(For this ${\bf Q}_0$, $k_{Fy}=k_{Fz}=0$)

\begin{equation}
\epsilon_{{\bf k}+{\bf Q}_0} = 
v_F p_x+\frac{p_y^2}{2m}+ \frac{p_z^2}{2m}+O(p_x^2)
\label{exp2}
\end{equation}

Now $E_{\bf k}$ and $E_{{\bf k}+{\bf Q}}$ can be expand as follows.
\begin{eqnarray}
E_{\bf k} &=& \Delta + \frac{\epsilon_{\bf k}^2}{2\Delta} \nonumber\\
E_{{\bf k}+{\bf Q}_0} &=& \Delta + 
\frac{\epsilon_{{\bf k}+{\bf Q}}^2}{2\Delta} 
\label{gap}
\end{eqnarray}
The integral of Eq. (\ref{int1}) becomes
\begin{equation}
\int_0^{\infty} d^d p
\frac{1}{(2\Delta -\omega) +\frac{1}{\Delta} [v_F^2 p_x^2+(p_y^2/2m+p_z^2/2m)^2]}
\end{equation}

One can see immediately by power counting,
for three dimensions, we have $log|\omega-2\Delta|$ singularity
as ${\bf p}$ appproachs $0$ (means ${\bf k} \rightarrow {\bf k}_F$).
For two dimensions,  we have $[\omega-2\Delta]^{-1/4}$ singularity.

\vskip 5mm

For general ${\bf Q}$, again choose ${\bf Q}$ parallel to $x$-axis
without loss of any generality and define
${\bf Q} =2 Q_x \hat{x} = 2k_{Fx} \hat{x}$ (See Fig. A2).
We again expand in terms of the $p$'s defined above.

\begin{eqnarray}
\epsilon_{\bf k} & = & -\frac{Q_x}{m}p_x +
  \frac{k_{Fy}}{m}p_y+\frac{k_{Fz}}{m}p_z
+\frac{p_x^2}{2m}+\frac{p_y^2}{2m}+ \frac{p_z^2}{2m} \nonumber\\
\epsilon_{{\bf k}+{\bf Q}} & =&  
\frac{Q_x}{m} p_x + \frac{k_{Fy}}{m}p_y+\frac{k_{Fz}}{m}p_z
+\frac{p_x^2}{2m}+\frac{p_y^2}{2m}+ \frac{p_z^2}{2m}
\label{exp3}
\end{eqnarray}

Unlike Eqs. (\ref{exp1}) and (\ref{exp2}), linear terms in $p_y$
 and $p_z$ are present in
Eq. (\ref{exp3}). This changes the power counting.
When we do the same expansion as in Eq. (\ref{gap})
 and do the integral, we see by just power counting that
in the lowest order of polarizability the $log$ is cut off by the
$k_{Fy}^2/2m$ and $k_{Fz}^2/2m$ terms in three dimensions.
(In fact, it has $log$ singularity in two dimensions
with coefficient depending on ${\bf Q}$ and changing to the stronger
singularity mentioned above at ${\bf Q}_0=2 k_F \hat{x} $)

To summarize,
singularity at $\omega =2\Delta$ for $d=3$ is found
for only  ${\bf Q}_0=2  k_F \hat{x}$ (or $<< \xi^{-1}$) where
the leading terms of $p_x$ is linear and $p_y$, $p_z$
are quadratic in Eq. (\ref{exp3}).

\section{}
We show here how to evaluate the integrals to get the results quoted in 
section IV.
We start with Eq. (\ref{charge}) in section II.
Expand $\epsilon_{\bf k}$ near the Fermi-surface and $\Delta_{\bf k}$ near
the maximum gap(See Eq. (\ref{model}) and (\ref{dgap}) in section II).
For ${\bf k}=(-k_F+p_x,p_y)$ 
\begin{equation}
E_{\bf k}\simeq \Delta-2\Delta p_y^2+\frac{E_F^2}{\Delta}
(2p_x^2-2p_x p_y^2+p_y^4/2),
\end{equation}
where $E_F=k_F^2/(2m)$ and 
$p_{x(y)}\equiv p_{x(y)}/k_F$ (we set $k_F=1$).
Note that it contains $2\Delta p_y^2$ term due to the anisotropy of the gap.
We can also find  $E_{{\bf k}+{\bf Q}_0}$ near ${\bf Q}_0$;
$(k_x,k_y)=(k_F-q_x+p_x,p_y+q_y)$.
The expression for $E_{\bf k}+E_{{\bf k}+{\bf Q}_0}-\omega$ 
is written as follows.
\begin{eqnarray}
f(p_x,p_y)&=&E_k+E_{k+Q}-\omega\nonumber\\
&=&\frac{E_F^2}{\Delta}(4p_x^2+4 p_x q_y p_y
+2 q_y^2 p_y^2+2 q_y p_y^3+p_y^4)-4\Delta q_y p_y-4\Delta p_y^2
+2\Delta-2\Delta q_y^2-\omega.
\end{eqnarray}
The susceptibility is obtained by integrating the $1/f(p_x,p_y)$ with respect
to $p_x$ and $p_y$.
After integrating over $p_x$ $(-\Delta/E_F < p_x < \Delta/E_F)$
we have 
\begin{eqnarray}
\int \frac{1}{f(p_x,p_y)} dp_x
&=&\frac{\Delta}{E_F^2} \frac{1}
{\sqrt{(p_y^2+q_y p_y-\frac{2\Delta^2}{E_F^2})^2+
\frac{\delta\omega \Delta}{E_F^2}-\frac{4\Delta^4}{E_F^4}}}\nonumber\\
&\times& \arctan{\frac{q_y p_y+2}
{\sqrt{(p_y^2+q_y p_y-\frac{2\Delta^2}{E_F^2})^2+
\frac{\delta\omega \Delta}{E_F^2}-\frac{4\Delta^4}{E_F^4}}}}
\end{eqnarray}
The leading term of this function is 
\begin{equation}
\frac{\pi}{2} \frac{\Delta}{E_F^2} \frac{1}
{\sqrt{(p_y^2+q_y p_y-\frac{2\Delta^2}{E_F^2})^2+
\frac{\delta\omega \Delta}{E_F^2}-\frac{4\Delta^4}{E_F^4}}}
\end{equation}

After performing the intergration over $p_y$,
we have the Elliptical function. 
\begin{equation}
K(k)=\int^1_0 \frac{dx}{\sqrt{(1-x^2)(1-k^2 x^2)}},
\end{equation}
where 
\begin{equation}
k=
 \left( \frac{
\sqrt{1+(E_F^2 q_y^2)/(8\Delta^2)-\sqrt{1-\delta\omega E^2_F/(4\Delta^3)}}}
{\sqrt{1+(E_F^2 q_y^2)/(8\Delta^2)+\sqrt{1-\delta\omega E^2_F/(4\Delta^3)}}}
\right). 
\end{equation}
When two poles of Elliptical function coincide( $k=1$), we have 
a $log$ singularity, which happens here at $\delta\omega = 4\Delta^3/
E_F^2$.
\begin{eqnarray}
{\rm Re} \chi_0({\bf Q}_0,\omega) &= & 
-\frac{1}{4\pi^2} \int \frac{1}{f(p_x,p_y)} dp_x dp_y
\nonumber\\
&=&
-\frac{N_0}{\sqrt{8+(E_F^2 q_y^2)/\Delta^2)}}
{\rm ln} \left( \frac{8\Delta}{|\delta\omega-4\Delta^3/E_F^2|}
\right),
\end{eqnarray}
where $N_0=1/(4\pi E_F)$.

The singular part of self-energy for spin-spin correlation function
is also presented here which are similar to those of the charge channel
apart from signs.
\begin{eqnarray}
{\rm Re}\Sigma_I &= & \frac{V^2N_0 r}{2\sqrt{2}} [
{\rm arcsin}{\frac{2\Delta}{\sqrt{E_F(3\Delta-\omega)}}}
\theta(-\omega+3\Delta-\frac{4\Delta^2}{E_F})\nonumber\\
& & -{\rm arcsin}{\frac{2\Delta}{\sqrt{E_F(3\Delta+\omega)}}}
\theta(\omega+3\Delta-\frac{4\Delta^2}{E_F})  ] \nonumber\\
{\rm Im}\Sigma_I &=& \frac{V^2N_0 r}{4\sqrt{2}} [
-{\rm arcsinh}{\frac{2\Delta}{\sqrt{E_F(-3\Delta+\omega)}}}
\theta(\omega-3\Delta)\nonumber\\
& & +{\rm arcsinh}{\frac{2\Delta}{\sqrt{E_F(-3\Delta-\omega)}}}
\theta(-\omega-3\Delta)  ]
\end{eqnarray}

\begin{eqnarray}
{\rm Re}\Sigma_{\tau_1} &= & \frac{V^2N_0 r}{2\sqrt{2}} [
{\rm arcsin}{\frac{2\Delta}{\sqrt{E_F(3\Delta-\omega)}}}
\theta(-\omega+3\Delta-\frac{4\Delta^2}{E_F})\nonumber\\
& & +{\rm arcsin}{\frac{2\Delta}{\sqrt{E_F(3\Delta+\omega)}}}
\theta(\omega+3\Delta-\frac{4\Delta^2}{E_F})  ] \nonumber\\
{\rm Im}\Sigma_{\tau_1} &=& -\frac{V^2N_0 r}{4\sqrt{2}} [
{\rm arcsinh}{\frac{2\Delta}{\sqrt{E_F(-3\Delta+\omega)}}}
\theta(\omega-3\Delta)\nonumber\\
& & +{\rm arcsinh}{\frac{2\Delta}{\sqrt{E_F(-3\Delta-\omega)}}}
\theta(-\omega-3\Delta)  ]
\end{eqnarray}
Note that the weight, $r$, here is smaller than that in 
Eqs. (\ref{selfre}) and (\ref{selfim}) by order of $\Delta/E_F$.

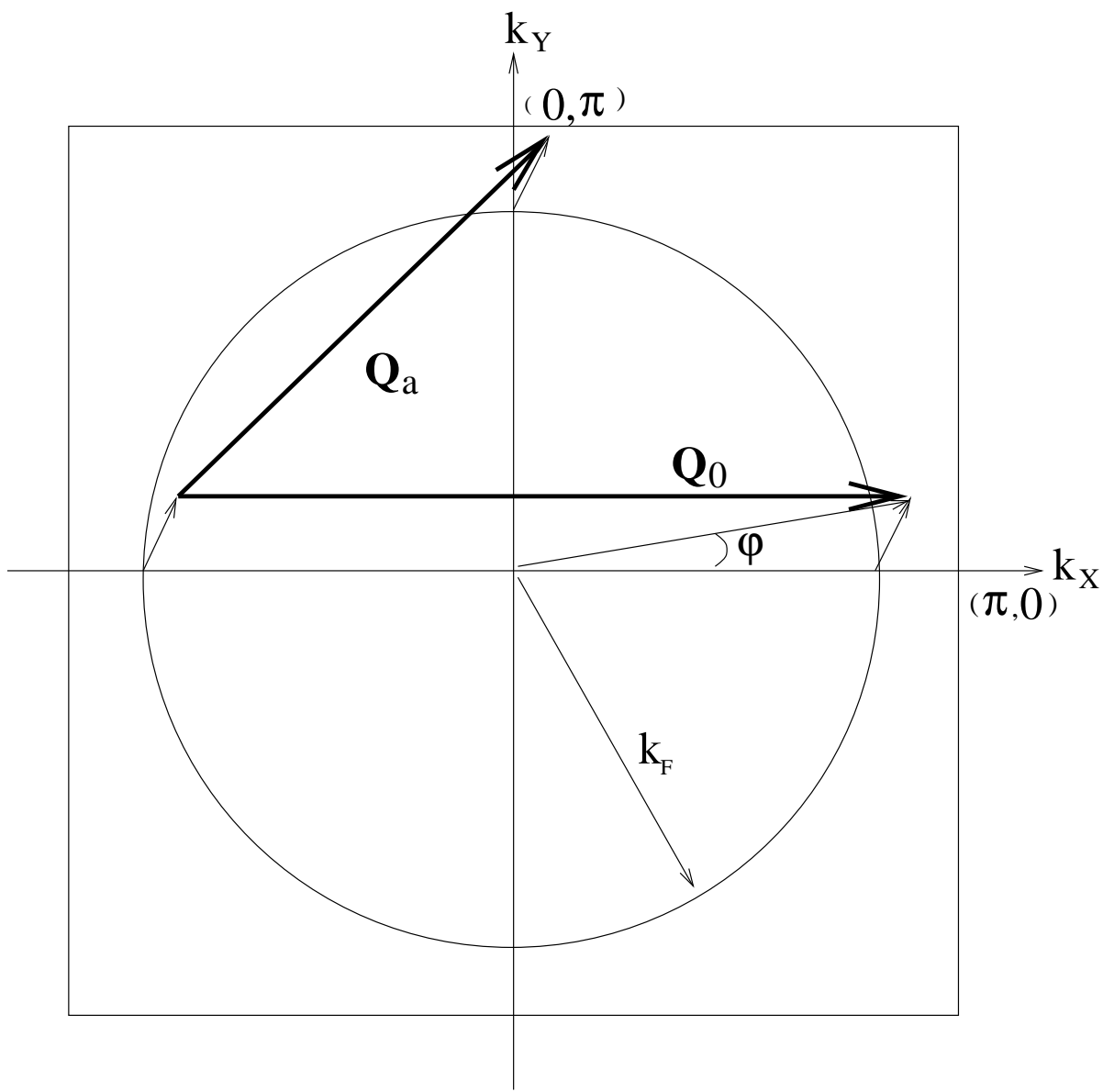
\begin{figure}
\caption{The isotropic Fermi-surface with radius $k_F$ illustrating
the vectors, 
${\bf Q}_0=(2 k_F,0)$ and ${\bf Q}_a=(k_F,k_F)$, used in the text.}
\end{figure}

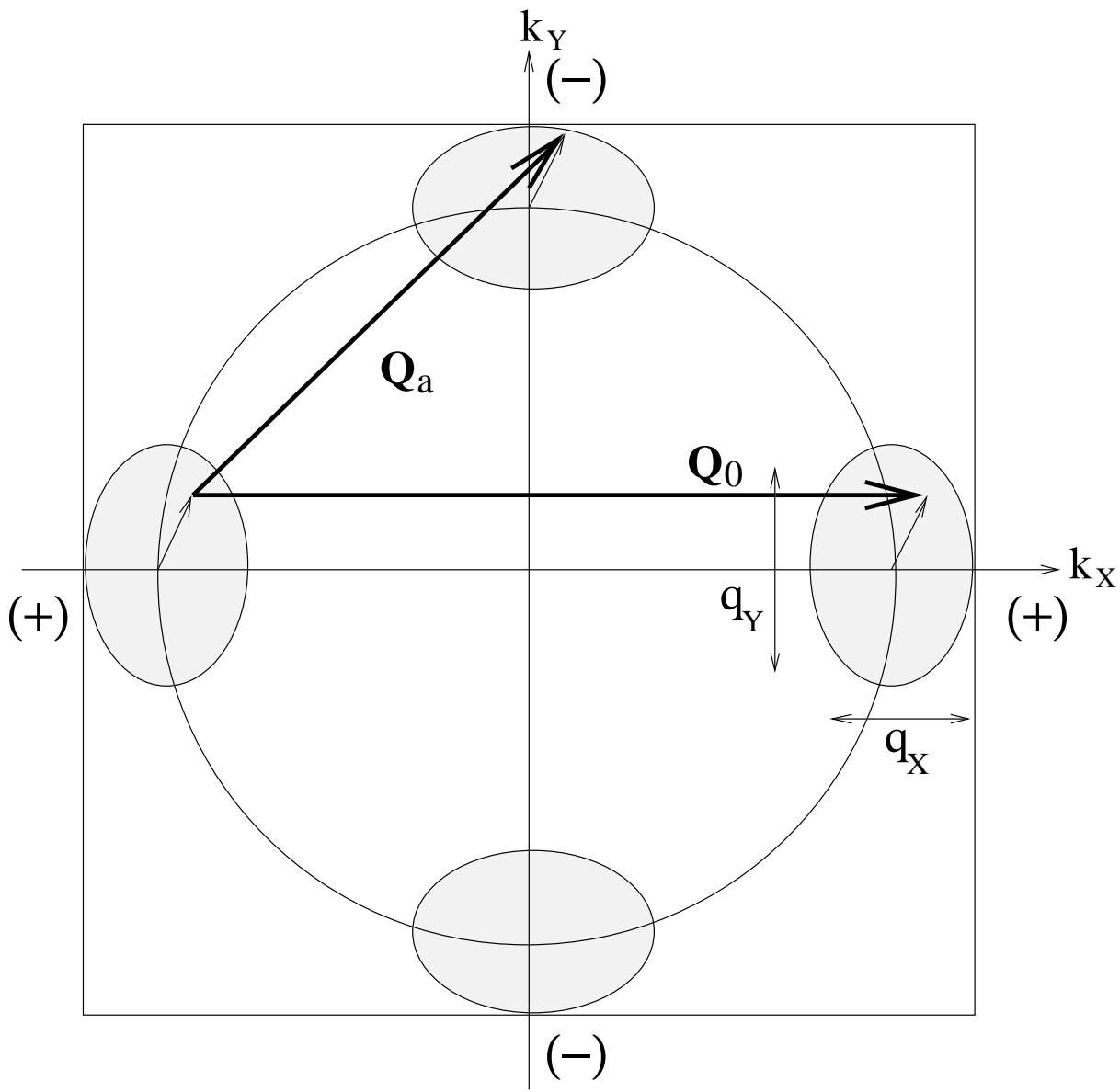
\begin{figure}
\caption{The isotropic Fermi-surface with d-wave 
gap symmetry illustrating vectors defined in the text. 
The $(+)$ and (-) denote the the signs of the gap. 
${\bf Q}_0$ and ${\bf Q}_a$ are momentum
transfer between two maximum gaps with same and  different signs,
respectively.
The range of $q_x$ and $q_y$ where our results for the polarizabilities
are valid are shown as shaded ellipse. 
}
\end{figure}

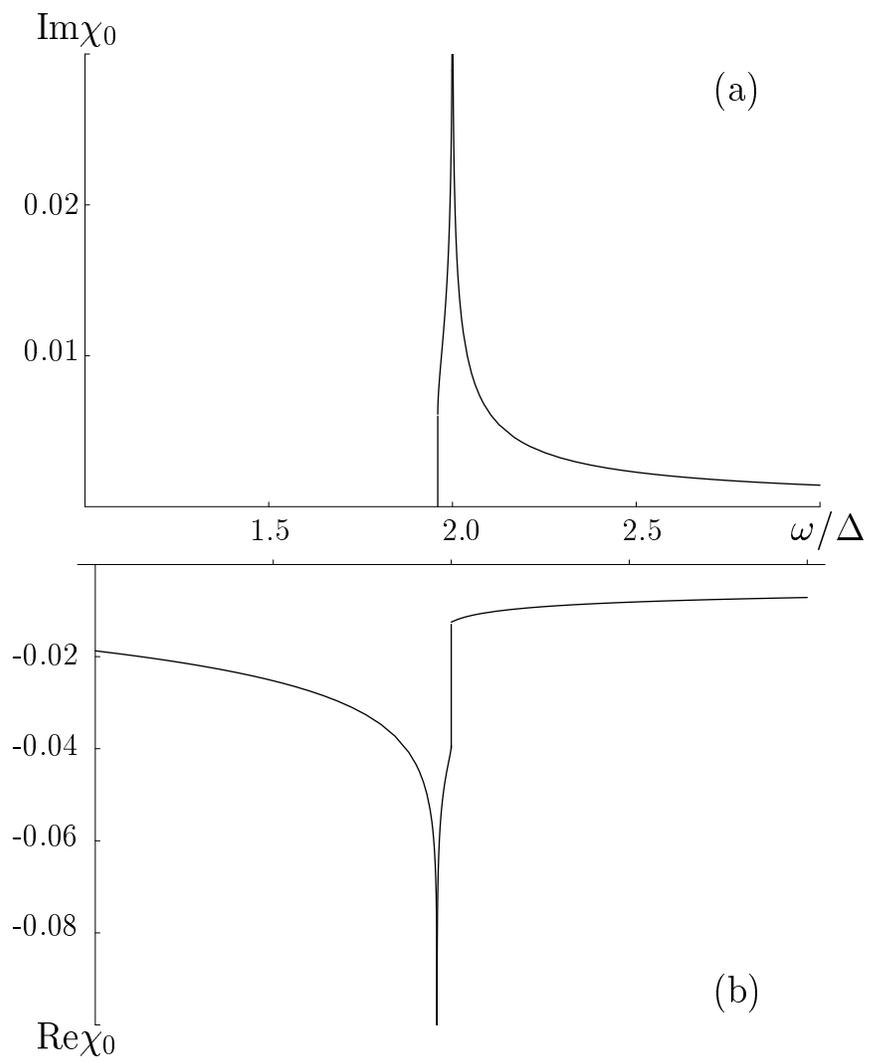
\begin{figure}
\caption{(a) The imaginary part of the lowest order polarizability,
${\rm Im} \chi^{charge}_0({\bf Q}_0,\omega)$. 
(b) The real part of the lowest order polarizability,
${\rm Re} \chi^{charge}_0({\bf Q}_0,\omega)$. 
The parameters are the same as those in
the text; $N_0 V \sim 1$ and $\Delta/E_F \sim \Delta/V \sim 0.1$.
}
\end{figure}

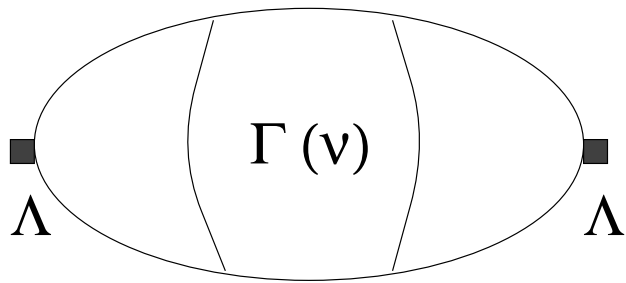
\begin{figure}
\caption{ The diagrammatic representation of the polarizability
with vertices $\Lambda$ and $\Gamma(\nu)$.}
\end{figure}

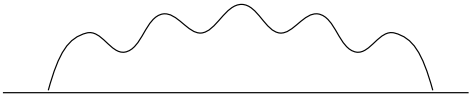
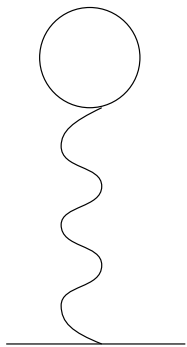
\begin{figure}
\caption{ (a) The renormalized interaction after RPA.
(b) The lowest order  self-energy; exchange contribution from
the Coulomb interaction
(c) The lowest order  self-energy; direct contribution from
the Coulomb  interaction.}
\end{figure}

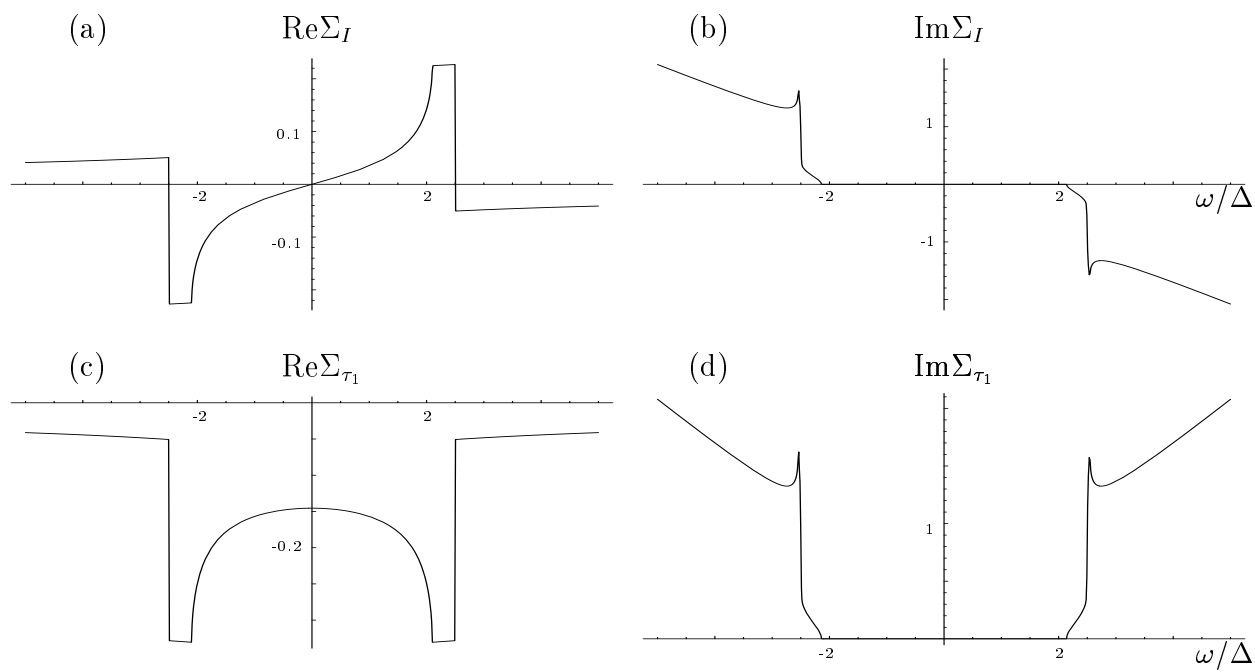
\begin{figure}
\caption{The lowest order self energy.
(a) ${\rm Re} \Sigma_I(\omega,{\bf k})$,
(b) ${\rm Im} \Sigma_I(\omega,{\bf k})$,
(c) ${\rm Re} \Sigma_{\tau_1}(\omega,{\bf k})$,
and (d) ${\rm Im} \Sigma_{\tau_1}(\omega,{\bf k})$,
where ${\bf k}$ is near the maximum gap.
The parameters are the same as those 
in the text; $N_0 V \sim 1$, $\Delta/E_F \sim \Delta/V \sim 0.1$,
and $r \sim 0.1$ }
\end{figure}

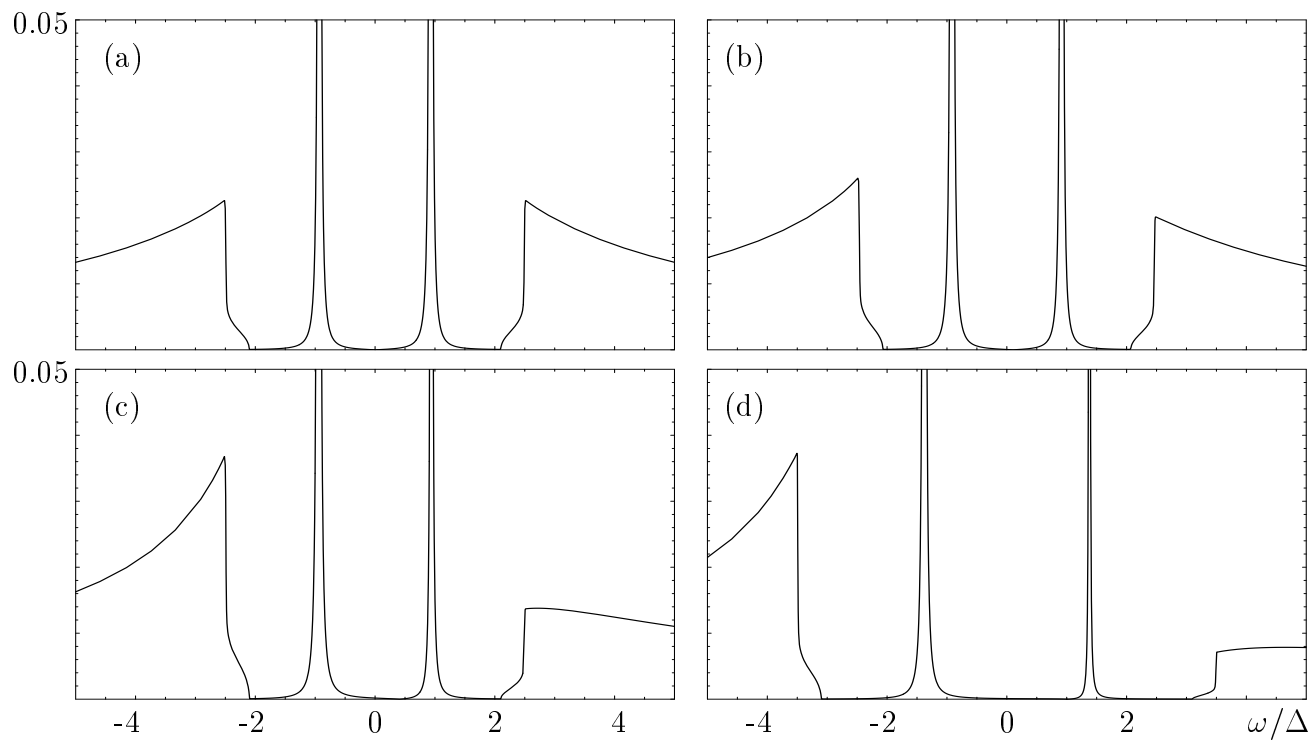
\begin{figure}
\caption{ The spectral function $A(\omega,{\bf k})$ with 
(a) ${\bf k} =( -k_F, 0)$,
(b) ${\bf k} =(-k_F, 0.1 k_F)$,
(c)${\bf k}=(-k_F, 0.2 k_F)$, and
(d) ${\bf k}=(-1.05 k_F,0)$. 
The parameters are the same as those 
in the text; $N_0 V \sim 1$, $\Delta/E_F \sim \Delta/V \sim 0.1$,
$r \sim 0.1$, and $r_{reg} \sim 0.9$}
\end{figure}

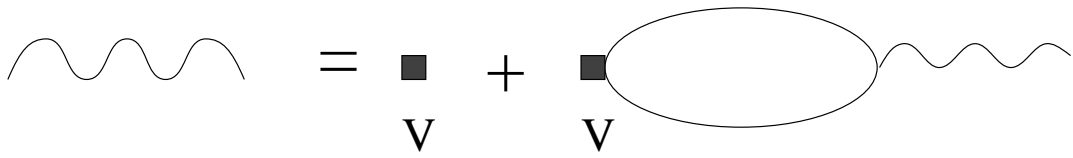
\end{document}